\documentclass{optica-article}

\journal{opticajournal} 

\articletype{Research Article}

\usepackage{lineno}

\begin{document}

\title{Linear and Nonlinear Coupling of Light in Twin-Resonators with Kerr Nonlinearity}

\author{Arghadeep Pal,\authormark{1,2,\dag} Alekhya Ghosh,\authormark{1,2,\dag}, Shuangyou Zhang\authormark{1}, Lewis Hill\authormark{1}, Haochen Yan\authormark{1,2}, Hao Zhang\authormark{1,3}, Toby Bi\authormark{1,2},Abdullah Alabbadi\authormark{1,2} and Pascal Del’Haye\authormark{1,2}}

\address{\authormark{1}Max Planck Institute for the Science of Light, 91058 Erlangen, Germany\\
\authormark{2}Department of Physics, Friedrich-Alexander-Universität Erlangen-Nürnberg, 91058 Erlangen, Germany\\
\authormark{3}National Key Laboratory of Microwave Photonics, Nanjing University of Aeronautics and Astronautics, Nanjing 210016, China\\
\authormark{\dag}The authors contributed equally to this work.\\}

\email{\authormark{*}pascal.delhaye@mpl.mpg.de} 


\begin{abstract*} 
Nonlinear effects in microresonators are efficient building blocks for all-optical computing and telecom systems. With the latest advances in microfabrication, coupled microresonators are used in a rapidly growing number of applications. In this work, we investigate the coupling between twin-resonators in the presence of Kerr-nonlinearity. We use an experimental setup with controllable coupling between two high-Q resonators and discuss the effects caused by the simultaneous presence of linear and non-linear coupling between the optical fields. Linear-coupling-induced mode splitting is observed at low input powers, with the controllable coupling leading to a tunable mode splitting. At high input powers, the hybridized resonances show spontaneous symmetry breaking (SSB) effects, in which the optical power is unevenly distributed between the resonators. Our experimental results are supported by a detailed theoretical model of nonlinear twin-resonators. With the recent interest in coupled resonator systems for neuromorphic computing, quantum systems, and optical frequency comb generation, our work provides important insights into the behavior of these systems at high circulating powers.

\end{abstract*}

\section{Introduction}
Microresonators are extremely well-suited for the direct observation of strong nonlinear interactions of light~\cite{kippenberg2011microresonator}. Small optical mode volumes and high quality ($Q$) factors enable low input threshold powers for observing interesting nonlinear effects in single resonators~\cite{shi2021thermal,guo2019nonclassical,ghadi2015two,chen2012bistability}. In particular, the Kerr effect in microresonators plays a pivotal role in recent works on spontaneous symmetry breaking (SSB)~\cite{wright1985,Woodley2018Universal,Hill2020Effects,delbino2017symmetry,Cao2017Experimental,woodley2021self,delbino2018microresonator,hill2024symmetry} and optical frequency combs~\cite{del2007optical}. Their applications~\cite{kippenberg2011microresonator} range from sensing~\cite{trocha2018ultrafast,yan2024real}, spectroscopy~\cite{yu2017microresonator}, telecommunications~\cite{alic2014frequency}, and waveform generation~\cite{zhang2019terahertz} to all-optical switching, photonic memories~\cite{delbino2021optical}, optically controllable circulators, isolators~\cite{white2023integrated} and logic gates~\cite{moroney2020logic}. 

Coupled resonator systems, in which the evanescent field of one resonator interacts with the optical mode in the adjacent one, have garnered much attention in recent years for demonstrating a myriad of new physical phenomena. Compared to a single resonator, these systems provide a rich platform for applications like slow light generation~\cite{melloni2003linear,chen2004nonlinearity,guo2021transition}, phonon lasing~\cite{grudinin2010phonon}, backscattering-induced higher-order filter design~\cite{hua2017high}, and optical sensing~\cite{fan2017tunable}. Parity-time symmetry has been experimentally shown using active-passive coupled microtoroids~\cite{wen2014pt}. These studies on coupled toroids discuss systems with controllable linear coupling between resonators, without exploring the high-power nonlinear regime and interactions of counterpropagating fields. Recent theoretical work predicts SSB in coupled ring resonators~\cite{ghosh2023four,ghosh2024controlled}. In addition, coupled Fabry-Pérot resonators with one linear and one nonlinear resonator~\cite{mai2022nonreciprocal,cheah2023spontaneous} have been studied. In the field of soliton frequency combs, a rich variety of fast-time dynamics (i.e., the time scale of one cavity round-trip) has been demonstrated experimentally using coupled microresonators~\cite{pidgaiko2023dispersion,miller2015tunable,komagata2021dissipative,tikan2021emergent} and theoretically in cavity-less two-component systems~\cite{sakaguchi2011symmetry,sakaguchi2021symmetry}. It is noteworthy that, the linear coupling between identical resonators causes the appearance of split-resonances, each of which is a linear superposition of symmetrical optical power distributions among the resonators (with different relative phases). However, SSB causes asymmetric optical power distribution among the coupled resonators.

In this work, we investigate a two-resonator system with tunable inter-resonator coupling and two coupling waveguides, where each resonator is pumped unidirectionally. Previous studies~\cite{Hill2020Effects,delbino2017symmetry} discuss symmetry breaking of light fields without considering inter-field linear scattering, whereas in~\cite{tikan2021emergent,komagata2021dissipative}, the coupling between the fields arises from the scattering of light from one resonator to the other, thereby not accounting for the Kerr effect induced symmetry broken homogeneous steady-state solutions. Our work considers both Kerr effect induced cross-phase modulation of the counterpropagating fields in each resonator, as well as linear scattering of light between the resonators. Systematic theoretical as well as experimental investigations reveal resonance splitting and shifting phenomena induced by the linear inter-resonator coupling and nonlinear Kerr interactions respectively. Here, we explore the slow-time (i.e., the time scale of multiple cavity round-trips) evolutions of fields in coupled high-Q microresonators at different input power levels with real-time controllable coupling between them. We also predict the occurrence of a novel nonlinear spontaneous symmetry breaking mechanism within each split-resonance of two coupled resonators. These effects significantly influence the optical power distribution within a coupled-resonator system. 

Section 2 introduces the governing equation that describes our coupled resonator system. The following sections focus on the experimental setup (Section 3.1) along with the interplay of the circulating fields in the two-resonator system at low power (Section 3.2) and at high power (Section 3.3). Both sections initially highlight the field dynamics in the twin resonators from a theoretical perspective with no inter-resonator coupling between them. Then the fabrication-induced asymmetries in the resonances (detuning, linewidth) of the two resonators, as observed in experiments, are considered to yield both theoretical and experimental results. Finally, the effect of increasing the inter-resonator coupling is presented in each section. 

This work is of importance for applications like real-time dispersion engineering, all-optical switching, neuromorphic computing, active light distribution in telecom systems, and optical sensing.

\begin{figure*}
\centering\includegraphics[width=1\columnwidth]{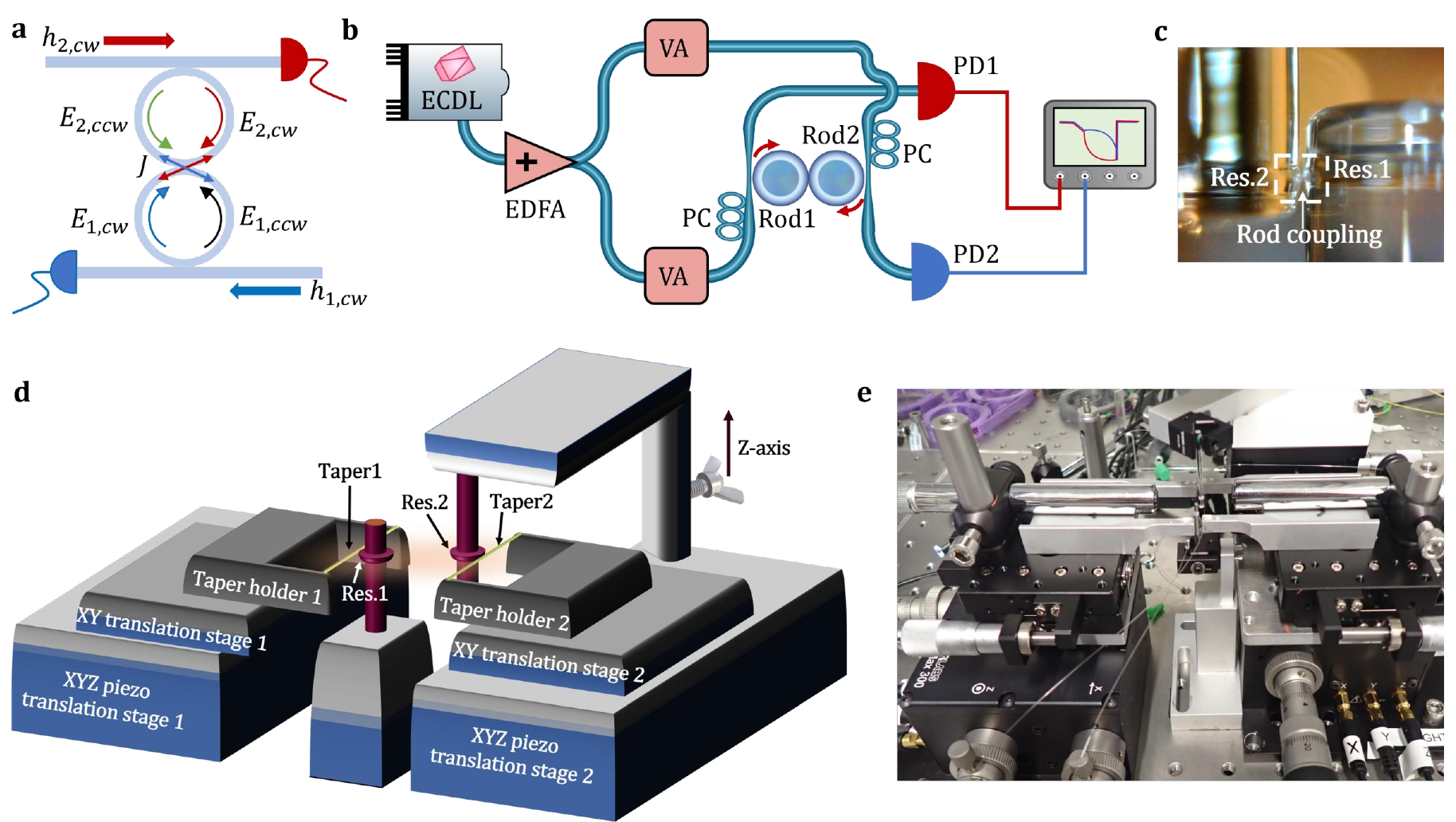}
\caption { Schematics and experimental setup. (a) Schematics showing the four fields ($E_{1,\text{cw}}$, 
 $E_{1,\text{ccw}}$, $E_{2,\text{cw}}$, $E_{2,\text{ccw}}$) circulating within the two coupled resonators with two inputs. (b) Experimental setup used to observe the linear and nonlinear interactions of light within the coupled twin-resonator system. ECDL: External cavity diode laser, EDFA: Erbium-doped fiber amplifier, VA: Variable attenuator, PD: Photodiode.(c) Side view photograph of the two coupled rod-resonators used in our experiments with the coupling region highlighted in the white dashed box. (d) Schematics showing the rod-taper coupling as well as the inter-resonator coupling mechanism used in our experimental setup. The fixed resonator is denoted as Res.1, whereas the invertedly clamped resonator is Res.2. (e) Image of the resonators, tapered fibers, and the translational stages used in our experiment.}
\label{fig1}
\end{figure*}

\section{Model}
The coupled Lugiato-Lefever equation (LLE) that describes the slow time evolution of the optical field envelopes takes the form~\cite{cheah2023spontaneous, komagata2021dissipative}, 
\begin{equation}
\frac{\partial E_{\ell,\pm}}{\partial\tau}=-\left(\frac{\kappa_\ell}{2}+i\delta\omega_{0,\ell}\right)E_{\ell,\pm}+iJE_{\ell^\prime,\mp}+i\left|E_{\ell,\pm}\right|^2E_{\ell,\pm}+i2\left|E_{\ell,\mp}\right|^2E_{\ell,\pm}+\sqrt{\kappa_{\text{ex},\ell}}h_{\ell,\pm},\label{LLEquations}
\end{equation}
where $E_{\ell,\pm}$ is the normalized optical field envelope in the $\ell^{\text{th}}$ resonator ($\ell$,$\ell^{'}$=$1,2$ and $\ell^\prime\neq\ell$) and can be written as $E_{\ell,\pm}=\sqrt{g_0}A_{\ell,\pm}$, where $g_{0}$ is the Kerr coefficient, $A_{\ell,\pm}$ is the unnormalized optical field envelope in the $\ell^{\text{th}}$ resonator circulating in a particular direction with “+” and “-” signs in the suffix representing the clockwise (CW) and counter-clockwise (CCW) propagating directions, respectively. The normalization using $g_0$ is performed such that to make the coupled LLEs independent of the Kerr-coefficient which is same for both the resonators. The coupling strength between the two resonators is $J$. The external coupling loss $\kappa_{\text{ex},\ell}$ constitutes the total loss $\kappa_\ell$, together with the intrinsic loss $\kappa_{0,\ell}$ in the $\ell^{\text{th}}$ resonator. The laser-cavity detuning is given by $\delta\omega_{0,\ell}$. Here, the input to the $\ell^{\text{th}}$ resonator is $h_{\ell,\pm}=\sqrt{g_0}s_{\ell,\pm}e^{-i\varphi_\ell}$, with $\ s_{\ell,\pm}$ and $\varphi_\ell$  being the input field amplitude and the phase (considered $0$ afterwards) for a certain circulating direction respectively. The third and fourth terms on the right-hand side of Eq.~\eqref{LLEquations} account for the nonlinear effects arising from self-phase and cross-phase modulation respectively. In the following sections, the ``+'' and ``-'' signs in the subscripts are replaced with cw and ccw to highlight the direction of light propagation. In the simulations, we neglect thermal effects, since they do not affect the asymmetry in propagating light fields. The two resonators are similar in terms of their diameter and quality factor. In Fig.~\ref{fig1}(a), it is shown that each resonator is pumped unidirectionally. Without coupling between the resonators, the inputs are chosen such that the fields are propagating in CW direction within each resonator, i.e., $h_{\ell,\text{ccw}}=0$. Seeding the clockwise (or counter-clockwise) fields in both resonators generates counter-propagating fields in each resonator in the coupled case, without additional interference effects. At the same time, cross-phase modulation leads to the observation of nonlinear symmetry breaking. Bringing the resonators closer results in an increasing value of $J$ and hence also increases coupling to the CCW fields in each resonator. 

\section{Results and discussions}
\subsection{Experimental setup}
Figure~\ref{fig1}(b) shows the experimental setup that is used to investigate the linear and nonlinear coupling of light in the twin-resonator system. Light from an external cavity diode laser (ECDL) at $1550$ nm is amplified via an erbium-doped fiber amplifier (EDFA). The light from the EDFA is split into two branches that couple to the evanescent fields of two glass rod resonators in CW directions via two tapered optical fibers. We focus on a single polarization mode by adjusting the input light with a polarization controller. To avoid complications, we couple into the resonances away from any mode-crossings. The rod resonators are made using $\text{CO}_2$ laser machining with a radius of around $0.9$ mm~\cite{del2013laser}. They have loaded $Q$ factors of around ${3\times10}^7$. 

\indent A detailed schematic of our experimental setup is shown in Fig.~\ref{fig1}(d). The experimental setup focuses on coupling between each of the rod resonators and a tapered fiber and the real-time control of the inter-resonator coupling. One of the resonators (Res.1) is kept fixed and coupled to a tapered fiber using the translational stages. The other set contains another inverted rod (Res.2) coupled to a tapered fiber. Both Res.2 and tapered fiber can be brought closer to Res.1 using the piezo translation stages.

\subsection{Resonance splitting at low power }
For the ideal symmetric case, the two resonators are considered identical, i.e. the system parameters in Eq.~\eqref{LLEquations} such as cavity loss (intrinsic loss $\kappa_0$, coupling loss $\kappa_{\text{ex}}$, and total loss $\kappa$), detuning ($\delta\omega_0$) are the same for both resonators. The input fields are also considered identical, i.e. $h_{1}=h_{2}=h$. Thus, the fields propagating in the resonators become equal, i.e. $E_{1}=E_{2}=E$. At sufficiently low power, the terms arising from nonlinear effects like self-phase and cross-phase modulation become negligible. At first, we consider the uncoupled system, i.e., the inter-resonator coupling term $J=0$. When the fields reach the stationary states in Eq.~\eqref{LLEquations}, they do not change over time, i.e. $\frac{\partial E}{\partial\tau}=0$. Thus, in steady state, the field amplitude can be written as, 
\begin{equation}
    \begin{split}
        E=\frac{\sqrt{\kappa_{\text{ex}}}h}{\left(\frac{\kappa}{2}+i\delta\omega_0\right)},
    \end{split}
    \label{Eq2}
\end{equation}
Eq.~\eqref{Eq2} represents the Lorentzian profile of cold cavity resonances as a function of laser frequency detuning $\delta\omega_{0,\ell}$  ($= \omega_0 - \omega_{\ell}$, where $\omega_0$ is the laser frequency and $\omega_{\ell}$ is the resonance frequency of the $\ell^{\text{th}}$ resonator at low pump power).

\begin{figure*}
\centering\includegraphics[width=1\columnwidth]{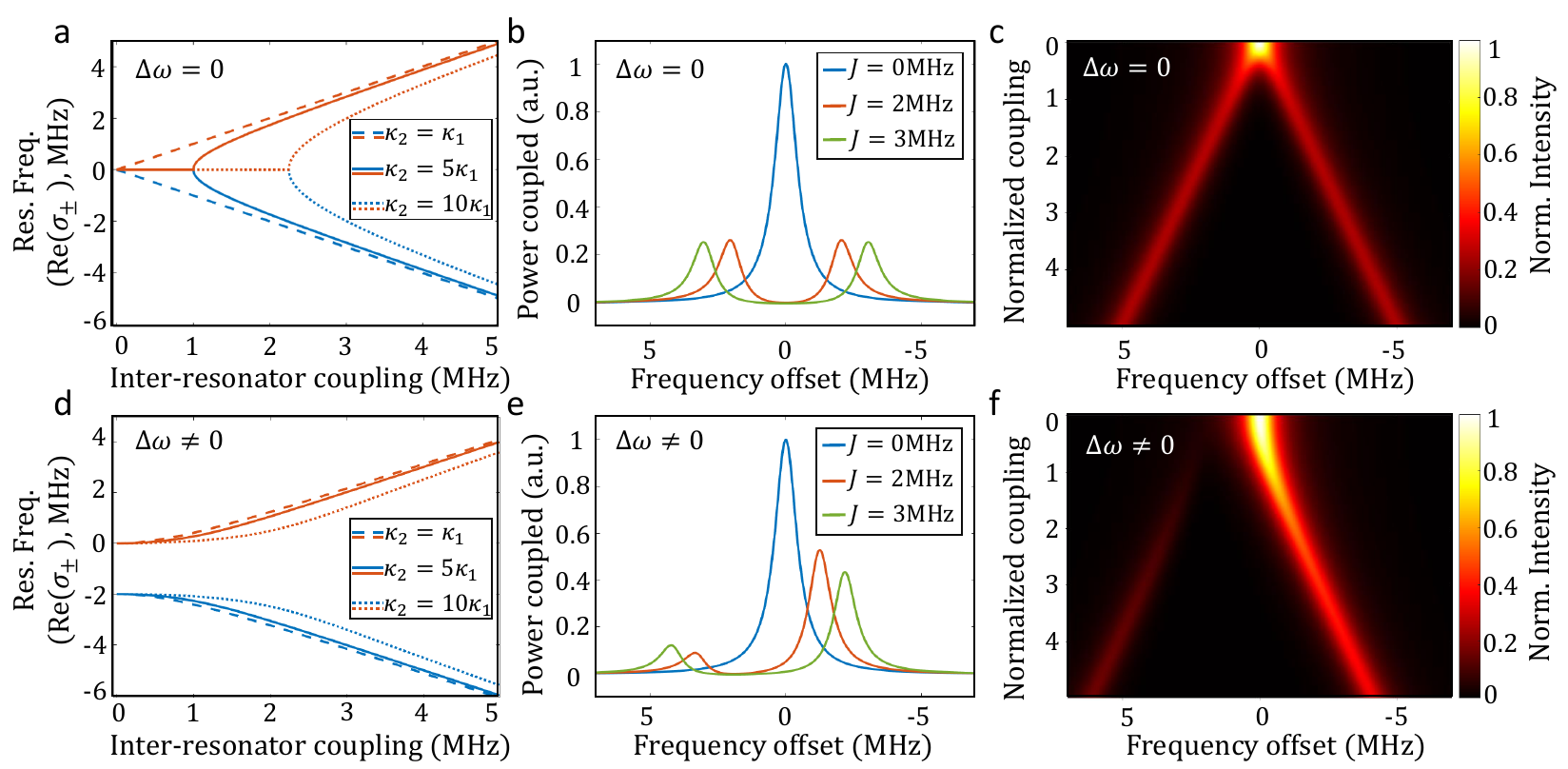}
\caption {Theoretical calculation of the resonance splitting in a coupled resonator system. Top (a, b, c) and bottom (d, e, f) panels correspond to coupled resonator systems with zero and nonzero differences in cold cavity resonance frequencies respectively. (a), (b) Eigenvalue analysis reveals resonance splitting dependence on inter-resonator coupling. The solid, dashed, and dotted lines show the resonance frequency splitting at different loss ratios of the two resonators. (b), (e) Power coupled into one of the resonators with respect to the frequency detuning. In each case, a new resonance appears for increasing values of the coupling $J$. (c), (f) Numerical analysis of Eq.~\eqref{LLEquations} shows an increasing resonance splitting with increasing inter-resonator coupling. Used parameter: $\kappa_{1}=$1~MHz.}
\label{fig2}
\end{figure*}

We first investigate the twin resonator system at low power. In the absence of inter-resonator coupling, the propagating optical fields are always circulating clockwise, however, as mentioned earlier, introduction of inter-resonator coupling initiates counter-clockwise light propagation in each resonator. This linear coupling induces symmetrical splitting of the resonances in both resonators. The field responses for the identical coupled resonator situation are presented in Fig.\ref{fig2}(a-c). In both experiments and simulations, the laser is scanned from high to low frequency. In order to inspect the splitting of the resonances at low power in more detail, we write Eq.~\eqref{LLEquations} in a matrix form and calculate the eigenvalues. Neglecting the nonlinear contributions, one can write Eq.~\eqref{LLEquations} in the following form,

\begin{equation}
    \begin{split}
        i\frac{d}{dt}\left[\begin{matrix}\begin{matrix}E_{1,\text{cw}}\\E_{1,\text{ccw}}\\\end{matrix}\\\begin{matrix}E_{2,\text{cw}}\\E_{2,\text{ccw}}\\\end{matrix}\\\end{matrix}\right]=\ M.\left[\begin{matrix}\begin{matrix}E_{1,\text{cw}}\\E_{1,\text{ccw}}\\\end{matrix}\\\begin{matrix}E_{2,\text{cw}}\\E_{2,\text{ccw}}\\\end{matrix}\\\end{matrix}\right]+\left[\begin{matrix}\begin{matrix}\sqrt{\kappa_{\text{ex},1}}h_{1,\text{cw}}\\0\\\end{matrix}\\\begin{matrix}\sqrt{\kappa_{\text{ex},2}}h_{2,\text{cw}}\\0\\\end{matrix}\\\end{matrix}\right],\ 
    \end{split}
    \label{Eq3}
\end{equation}

\noindent where, 

\begin{equation}
M=\left[\begin{matrix}-i\frac{\kappa_1}{2}+\delta\omega_{0,1}&0&0&-J\\0&-i\frac{\kappa_1}{2}+\delta\omega_{0,1}&-J&0\\0&-J&-i\frac{\kappa_2}{2}+\delta\omega_{0,2}&0\\-J&0&0&-i\frac{\kappa_2}{2}+\delta\omega_{0,2}\\\end{matrix}\right].
\end{equation}

The eigenvalues of the matrix $M$ are
\begin{equation}
    \begin{split}
        \sigma_\pm=-i\frac{\Delta\kappa}{4}+\frac{\Delta\omega}{2}+\delta\omega_{0,1}-i\frac{\kappa_1}{2}\pm\sqrt{\left(\frac{\Delta\omega}{2}\right)^2+J^2-\left(\frac{\Delta\kappa}{4}\right)^2-i\frac{\Delta\kappa.\Delta\omega}{4}}, 
    \end{split}
    \label{Eq4}
\end{equation}
\noindent where, $\Delta\kappa=\kappa_2-\kappa_1,\ \ \Delta\omega=\delta\omega_{0,2}-\delta\omega_{0,1}$. We have considered $\Delta\omega$ to be positive. The real part of the eigenvalues represents the hybridized mode frequencies.

When the two resonators are identical, the eigenvalues can be written as,  
\begin{equation}
    \begin{split}
        \sigma_\pm=\delta\omega_{0,1}-i\frac{\kappa_1}{2}\pm\sqrt{J^2}=-i\frac{\kappa_1}{2}+\delta\omega_{0,1}\pm J. 
    \end{split}
    \label{Eq5}
\end{equation}

Considering the system is at resonance, i.e., $\delta\omega_{0,1}=0$, $Re\left[\sigma_\pm\right]=\pm J$. Therefore, even an infinitesimal value of inter-resonator coupling causes a splitting of the resonance.

Now, we consider the situation when the two resonators have equal resonance frequencies, but different losses. In this case, considering the system being at resonance corresponds to $\sigma_\pm=-i\frac{\Delta\kappa}{4}-i\frac{\kappa_1}{2}\pm\sqrt{J^2-\left(\frac{\Delta\kappa}{4}\right)^2}$. Thus, for $\left(\Delta\kappa\right)^2>16J^2$, there is no resonance splitting. The resonance splits at the point $\left(\Delta\kappa\right)^2=16J^2$ and the amount of the splitting increases with coupling. This situation is shown in Fig.~\ref{fig2}(a-c). However, in real-world experiments, the resonance frequencies are never exactly identical. With increasing inter-resonator coupling, the difference between the resonance frequencies starts to grow significantly. The completely asymmetrical situation is depicted in Fig.~\ref{fig2}(d-f).

In reality, it is challenging to fabricate two identical resonators; thus, they have different resonance frequencies (resulting in different detunings from the laser frequency), and cavity losses. The high $Q$ factors and narrow resonance linewidths make the overlapping of resonance frequencies of two different resonators highly unlikely at low input powers without additional means to fine-tune one of the resonators. In the case of integrated systems, external heaters have been used~\cite{miller2015tunable} to align the resonance frequencies of the two resonators which are then coupled to achieve resonance splitting. However, in this work, we study the field dynamics in a coupled resonator system without manipulation of resonator properties via external control. Considering the asymmetry in the system and including the inter-resonator coupling term $J$, the CW circulating fields in Eq.~\eqref{LLEquations} can be rewritten as,
\begin{equation}
    \begin{split}
        E_{\ell,\text{cw}}=\frac{-\ \left(\frac{\kappa_\ell}{2}+i\delta\omega_{0,\ell}\right)\sqrt{\frac{\kappa_{\text{ex},\ell}}{8}}h_{\ell,\text{cw}}}{\left\{-\left(\frac{\kappa_\ell}{2}+i\delta\omega_{0,\ell}\right)\left(\frac{\kappa_\ell}{2}+i\delta\omega_{0,\ell}\right)-J^2\right\}}.\  
    \end{split}
    \label{Eq6}
\end{equation}

\begin{figure*}[t]
\centering\includegraphics[width=1\columnwidth]{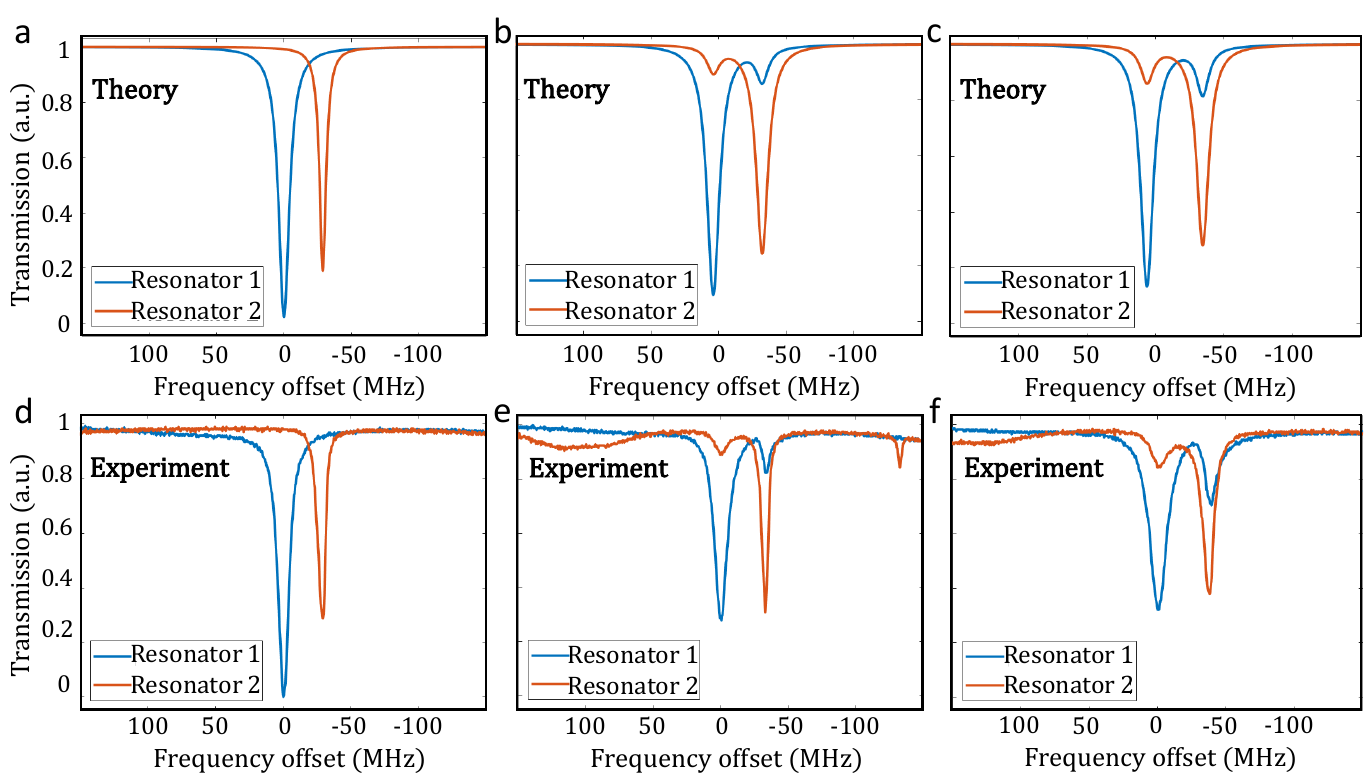}
\caption {Mode hybridization at different resonator coupling strengths. The top row shows theoretical results while the bottom row displays experimental data. The graphs are transmission spectra of the two resonators, plotted as a function of the laser frequency detuning from the cold resonance (the resonance of an uncoupled resonator at low optical powers) of resonator 1. (a-c) Theoretically calculated transmission profiles for different values of $J$, while keeping other parameters of Eq.~\eqref{Eq3} constant ($\kappa_1=11.3$ MHz, $\kappa_{\text{ex},1}=5.96$ MHz, $\kappa_2=10.9$ MHz, $\kappa_{\text{ex},2}=4.71$ MHz, $\delta\omega_{0,1}-\delta\omega_{0,2}=2.5\kappa_1, h_{1,\text{cw}}=h_{2,\text{cw}}=3.4\kappa_{0,1}$). For the simulated scans, $J=0$ (a), $J=\kappa_1$ (b), $J=1.3\kappa_1$ (c). (d) Experimentally obtained transmission spectra when the two resonators are uncoupled. (e-f) Resonance splitting appears with increasing inter-resonator coupling.}
\label{fig3}
\end{figure*}

The theoretical calculation of the transmission spectrum as a function of the laser-cavity detuning is shown in Fig.~\ref{fig3}(a) for $J=0$ in Eq.~\eqref{Eq3}. Similar to the theory, the experimental result in Fig.~\ref{fig3}(d) depicts two non-overlapping Lorentzian profiles with different resonance frequencies and $Q$ factors. 

The effects of increasing inter-resonator coupling are simulated using Eq.~\eqref{Eq3} and the results are shown in Fig.~\ref{fig3}(b-c). Inter-resonator coupling results from the overlapping of the evanescent field of one resonator with the cavity mode within the other. Therefore, in experiments, the inter-resonator coupling can be increased by bringing the resonators closer, which increases the overlap of the fields. The experimental observations presented in Fig.~\ref{fig3}(e-f) are in good agreement with our theoretical prediction. An increase in the inter-resonator coupling causes the emergence of the hybridized modes. A change in inter-resonator coupling leads to a shift in the position of the hybridized resonances. Thus, this real-time coupling control is highly beneficial, e.g., for direct dispersion control of the resonators. By changing the inter-resonator coupling, we can adjust the relative resonance frequency positions as well as the circulating powers.

\subsection{Interplay of linear coupling and Kerr-nonlinearity at high input powers}
\begin{figure*}
\centering\includegraphics[width=1\columnwidth]{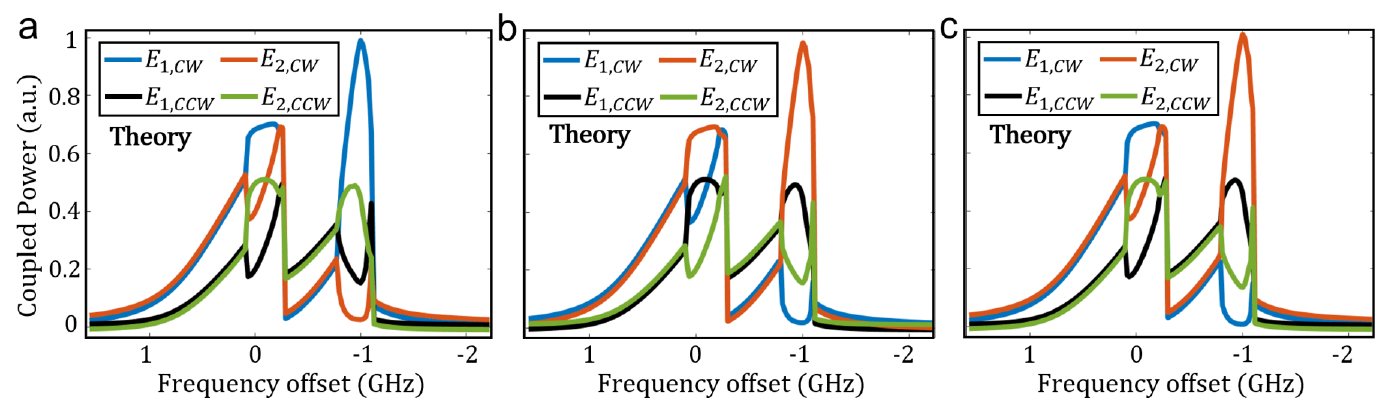}
\caption {Circulating field intensities at high power in identical twin-resonators while the input pump scans from high to low frequency at fixed power. (a-c) High circulating powers lead to symmetry breakings that result in sudden changes in the power distribution within the resonator modes. The three panels show different combinations of dominant modes after the symmetry breaking bifurcations. For better visualization, we added a tiny vertical offset to some of the colored graphs in order to avoid them being plotted on top of each other in the symmetric regions.}
\label{fig4}
\end{figure*}

At low pump power, the optical nonlinear effects are negligible, which leads to  linear scattering induced resonance splittings. With increasing input power, the optical nonlinearities start to influence the optical modes. Investigations at high input power reveal a new type of Kerr-nonlinearity-induced symmetry breaking of the different optical modes. This symmetry breaking leads to sudden differences in previously equal resonance frequencies. As a result, this induces a power redistribution between the modes of the two resonators, as shown in Fig.~\ref{fig4}. For clarification: a linear coupling between two resonator modes leads to a symmetric splitting of the resonances in each of the modes. The nonlinear symmetry breaking leads to a sudden resonance frequency difference of previously degenerate modes and not a mode splitting. In this section, we will delve into the theoretical details of the mechanism and present the experimental findings.

At higher input powers, the nonlinear terms in Eq.~\eqref{LLEquations} need to be considered. However, for $J=0$, the field propagates in the resonator in one direction and thus, the cross-phase modulation term is not present. For high input powers, it is not straightforward to find the homogeneous solution states of the system analytically and instead, we investigate the system numerically. While sweeping the input pump laser from a higher frequency towards the resonance, the Lorentzian seen at low input power in Fig.~\ref{fig2}(a) broadens at higher power to take the shape of a triangle that jumps out of resonance abruptly~\cite{carmon2004dynamical}.

In a symmetric uncoupled system, equal powers circulate in the two microresonators in clockwise direction. When the two resonators are coupled to each other, the clockwise propagating field of each resonator couples to the counterclockwise mode of the other resonator. As mentioned earlier, the resonances split depending on the values of the inter-resonator coupling. For high input powers, both split-resonances are starting to be skewed and assume a triangular shape. This phenomenon is presented in Fig.~\ref{fig4}. The parameters used in the simulations are $\kappa_{0,1}=\kappa_{0,2}=\kappa_{\text{ex},1}=\kappa_{\text{ex},2}=157.08$ MHz, $J=3.4\kappa_{0,1}$, and $h_{1,\text{cw}}=h_{2,\text{cw}}=4.07\kappa_{0,1}$. The power is kept constant and the pump frequency is tuned from higher to lower values at around $7.3~$GHz/sec.

\begin{figure*}
\centering\includegraphics[width=1\columnwidth]{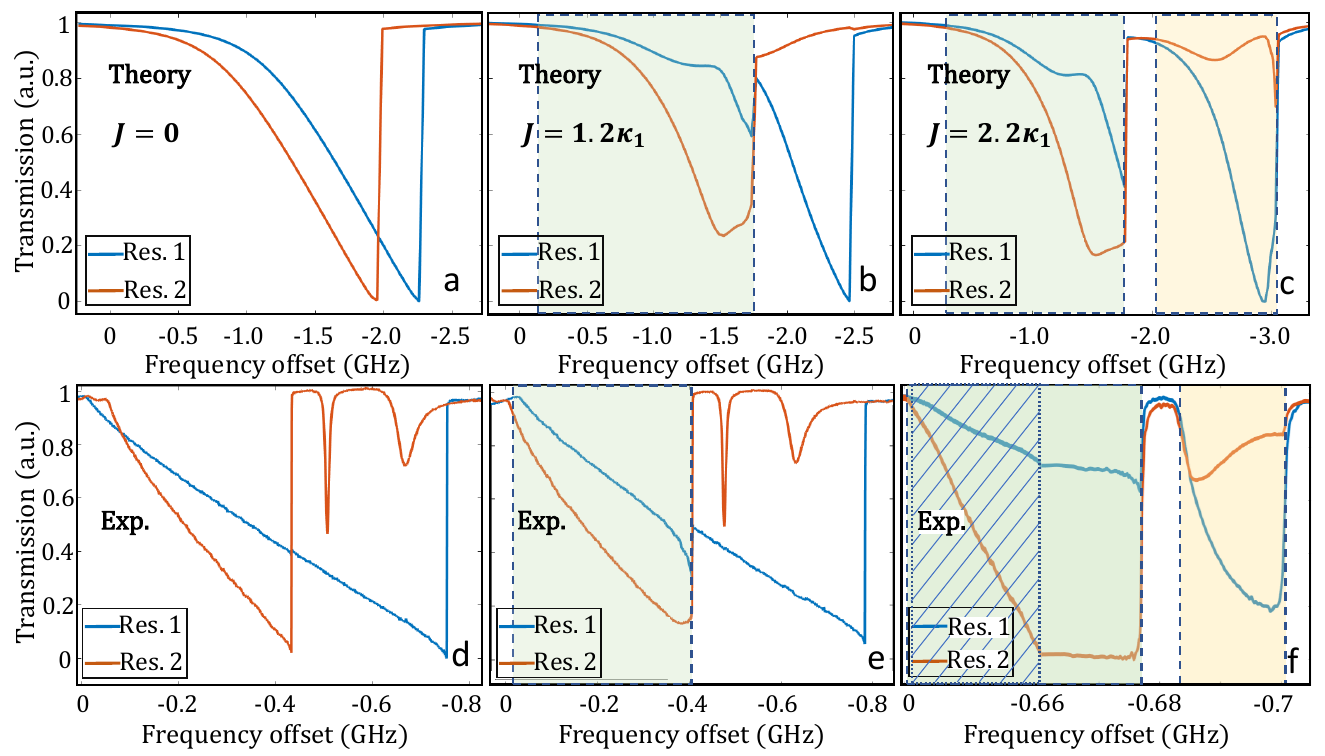}
\caption {Resonance spectra of two asymmetric microresonators at high input power. (a-c) Simulated transmission profiles as a function of the laser frequency offset are plotted with varying $J$ while other parameters of Eq.~\eqref{LLEquations} are constant ($\kappa_{12}={2\kappa}_{\text{ex},1,2}=314$MHz, $\delta\omega_{0,1}-\delta\omega_{0,2}=\kappa_1,\ h_{1,\text{cw}}=h_{2,\text{cw}}={3.4\kappa}_1$ ). The different values of inter-resonator coupling strength $J$ are (a) $0$, (b) $1.2\kappa_1$, (c) $2.2\kappa_1$. (d-f) Transmission plots of the resonators while increasing the inter-resonator coupling from uncoupled (d) to maximum inter-resonator coupling (f), where the long monotonous region is horizontally compressed in the shaded region. The green region in (b), (c), (e), and (f) shows the overlap of the first split-resonances (higher frequency ones) of the two resonators, where the yellow shaded region in (c) and (f) highlights the field interactions of the split resonance at lower frequencies.}
\label{fig5}
\end{figure*}
At high powers, even if equal input powers are launched into the resonators, self-amplification of small power fluctuations between the input fields (clockwise fields) leads to a power imbalance~\cite{delbino2017symmetry}. This initiates the breaking of the symmetry of the clockwise fields in the two resonators and the asymmetry is enhanced by Kerr nonlinearity. An opposite trend of dominance for the counter-clockwise fields in the two resonators is observed. This is because, the more the clockwise field in one resonator, the more light gets coupled to the other resonator in counter-clockwise direction.
Figure~\ref{fig4} shows the simulations of the powers circulating in different directions in the two resonators with the occurrence of a unique symmetry breaking phenomenon. The clockwise inputs provided in the system are indicated by blue (resonator 1 in Fig.~\ref{fig1}(a)) and red (resonator 2 in Fig.~\ref{fig1}(a)) with the black and green lines being their respective counter-propagating fields. The amplitudes of counter-clockwise fields depend on the coupling between the resonators. In the first split-resonance (higher frequency), beyond a certain detuning threshold, the circulating clockwise power of one resonator (blue) increases in comparison to that of the second resonator (red) (Fig.~\ref{fig4}(a)). In the perfectly symmetric case (identical resonators), the dominant mode is chosen randomly. The stronger field (blue) of the resonator pushes its counter-circulating (black) mode’s resonance away from the cold resonance position and also from the pump laser due to the Kerr-shift induced by cross-phase modulation. Thus, less power couples into the field direction indicated by the black line. We can see opposite changes in the circulating powers in the two directions (blue and black in Fig.~\ref{fig4}). Since the clockwise direction of resonator 1 in Fig.~\ref{fig1}(a) dominates in this case (Fig.~\ref{fig4}(a)), the coupled power to the counter-clockwise direction in the other resonator goes up. This increase pushes the clockwise resonance of resonator 2 in Fig.~\ref{fig1}(a) away from the cold resonance position due to cross-phase modulation and causes a decrease in the circulating power. This new type of symmetry breaking is a result of the interplay between both nonlinear Kerr-effect and linear coupling between the resonators. Symmetry breaking phenomena have been also observed in the second split-resonance (the split-resonance mode that has a relatively higher resonance wavelength, i.e. lower frequency). Since random input power fluctuations lead to this effect, the dominant modes are chosen randomly in each of the split resonances and thus Fig.~\ref{fig4} (b) and (c) show the changing of the fields’ dominance in each of the split resonance modes. In actual experiments, it is not possible to obtain a perfectly symmetric pair of resonators. Thus, we investigate the linear and nonlinear effects in slightly asymmetric coupled resonator systems. Figure~\ref{fig5} depicts a comparison between our theoretical calculations using Eq.~\eqref{LLEquations} (top panel) and experimental results (bottom panel) of the transmission profiles of two asymmetric resonators with varying inter-resonator coupling at high power. For the uncoupled system ($J=0$), the triangular-shaped resonance dips in the transmission spectra of two asymmetric resonators obtained from numerical simulations of Eq.~\eqref{LLEquations} are shown in Fig.~\ref{fig5}(a), which has been replicated experimentally in Fig.~\ref{fig5}(d). In the high input power regime, the Kerr effect causes an overlap of the resonances of two non-identical resonators. This assures each resonator has sufficient circulating power in the overlapping frequency range to drive nonlinear effects. Therefore, we observe Kerr-nonlinear effects along with linear coupling induced resonance splitting. Thermal effects symmetrically influence all the fields in the resonator unlike direction dependent Kerr-nonlinear effects (i.e., the strengths of cross-phase modulation and self-phase modulations being different). Thus, we neglect thermally induced resonance shifts in our theoretical calculations.

For asymmetric resonators (different losses) and increasing coupling, we first observe the splitting of the high-Q (CW resonance in resonator 1). In this case, the low-Q resonance (CCW resonance in resonator 2) spectrally overlaps with the high-frequency part of the split-resonance of resonator 1 (green shaded regions in Figs.~\ref{fig5}(b,e)), demonstrating nonlinear enhancements of asymmetries of circulating field intensities in the two resonators~\cite{garbin2020asymmetric}. With further increase in coupling, the low-Q resonance splits as well. Two frequency regions arise, in which the two split-resonances of the resonators overlap (green and yellow regions in Figs.~\ref{fig5}(c,f)). The nonlinear enhancements of asymmetries are observed within both overlapping regions simultaneously.

Both theoretical studies and experimental results conclude that at high power, an increase in the inter-resonator coupling strength leads to the observation of both linear and non-linear interactions simultaneously. Note that the experimental results in Fig.~\ref{fig5}(d-f) show the presence of some very low Q resonances in resonator 1 (red) which do not have any impact on the nonlinear interaction between the fields of two resonators. For very high input powers the modes will eventually show nonlinear broadening, however, here we focus on the overlapping regions of the higher Q resonances.

\section{Conclusions}
We investigate a system of twin-microresonators with both linear and Kerr-nonlinear coupling. At low input powers, the emergence of new hybrid resonances is observed. Increasing the coupling between the two resonators leads to an increasing frequency splitting between the hybrid resonances. At higher input powers, the resonance splitting is accompanied by Kerr-nonlinearity-induced symmetry breaking that leads to asymmetric power distribution between the two resonators. Our theoretical calculations are in good agreement with the measured resonance spectra.

In addition to the fascinating physics, the nonlinear coupling between twin-resonators has many interesting applications. Real-time control of the inter-resonator coupling can for example enable precise dispersion control for soliton frequency comb generation. Unlike the contemporary dispersion engineering techniques~\cite{pal2023machine}, this instantaneous control is of interest for spectrally tailored soliton generation~\cite{helgason2021dissipative} and soliton pair generation~\cite{yuan2023soliton}. Additional applications include photonic data storage~\cite{zhang2019electronically}, and controlled lasing~\cite{peng2014loss}. In the future, this work can be extended to study systems of more coupled resonators for neuromorphic computing and optical data processing using integrated photonic circuits.

\begin{backmatter}
\bmsection{Funding}
This work was supported by the European Union’s H2020 ERC Starting Grant "CounterLight" 756966 and the Max Planck Society. AP and AG acknowledge the support from Max Planck School of Photonics. LH acknowledges funding provided by the SALTO funding scheme from the Max-Planck-Gesellschaft (MPG) and Centre national de la recherche scientifique (CNRS). 

\bmsection{Acknowledgments}
PDH, AP, and AG defined the project. AP and AG contributed equally to the work. AP led the experimental works with support from SZ and AG. AG led the theoretical calculations with support from LH and AP. SZ helped with the supervision of the experimental part of the project. LH helped with the supervision of the theory part of the project. AP, AG, and SZ wrote the manuscript with contributions from all the authors. 

\bmsection{Disclosures}
The authors declare no conflicts of interest.

\bmsection{Data Availability Statement}
A Data Availability Statement (DAS) will be required for all submissions beginning 1 March 2021. The DAS should be an unnumbered separate section titled ``Data Availability'' that
immediately follows the Disclosures section. See the \href{https://opg.optica.org/submit/review/data-availability-policy.cfm}{Data Availability Statement policy page} for more information.

\end{backmatter}




\bibliography{sample}

\begin{thebibliography}{10}
\newcommand{\enquote}[1]{``#1''}

\bibitem{kippenberg2011microresonator}
T.~J. Kippenberg, R.~Holzwarth, and S.~A. Diddams, \enquote{Microresonator-based optical frequency combs,} {\protect\JournalTitle{science}} \textbf{332}, 555--559 (2011).

\bibitem{shi2021thermal}
X.~Shi, W.~Fan, A.~K. Hansen, \emph{et~al.}, \enquote{Thermal behaviors and optical parametric oscillation in 4h-silicon carbide integrated platforms,} {\protect\JournalTitle{Advanced Photonics Research}} \textbf{2}, 2100068 (2021).

\bibitem{guo2019nonclassical}
K.~Guo, L.~Yang, X.~Shi, \emph{et~al.}, \enquote{Nonclassical optical bistability and resonance-locked regime of photon-pair sources using silicon microring resonator,} {\protect\JournalTitle{Physical Review Applied}} \textbf{11}, 034007 (2019).

\bibitem{ghadi2015two}
A.~Ghadi and S.~Mirzanejhad, \enquote{Two-photon absorption effect on semiconductor microring resonators,} {\protect\JournalTitle{Optik}} \textbf{126}, 1645--1649 (2015).

\bibitem{chen2012bistability}
S.~Chen, L.~Zhang, Y.~Fei, and T.~Cao, \enquote{Bistability and self-pulsation phenomena in silicon microring resonators based on nonlinear optical effects,} {\protect\JournalTitle{Optics Express}} \textbf{20}, 7454--7468 (2012).

\bibitem{wright1985}
E.~M. Wright, P.~Meystre, W.~J. Firth, and A.~E. Kaplan, \enquote{Theory of the nonlinear sagnac effect in a fiber-optic gyroscope,} {\protect\JournalTitle{Phys. Rev. A}} \textbf{32}, 2857--2863 (1985).

\bibitem{Woodley2018Universal}
M.~T.~M. Woodley, J.~M. Silver, L.~Hill, \emph{et~al.}, \enquote{Universal symmetry-breaking dynamics for the kerr interaction of counterpropagating light in dielectric ring resonators,} {\protect\JournalTitle{Phys. Rev. A}} \textbf{98}, 053863 (2018).

\bibitem{Hill2020Effects}
L.~Hill, G.-L. Oppo, M.~T.~M. Woodley, and P.~Del'Haye, \enquote{Effects of self- and cross-phase modulation on the spontaneous symmetry breaking of light in ring resonators,} {\protect\JournalTitle{Phys. Rev. A}} \textbf{101}, 013823 (2020).

\bibitem{delbino2017symmetry}
L.~Del~Bino, J.~M. Silver, S.~L. Stebbings, and P.~Del'Haye, \enquote{Symmetry breaking of counter-propagating light in a nonlinear resonator,} {\protect\JournalTitle{Scientific Reports}} \textbf{7}, 43142 (2017).

\bibitem{Cao2017Experimental}
Q.-T. Cao, H.~Wang, C.-H. Dong, \emph{et~al.}, \enquote{Experimental demonstration of spontaneous chirality in a nonlinear microresonator,} {\protect\JournalTitle{Phys. Rev. Lett.}} \textbf{118}, 033901 (2017).

\bibitem{woodley2021self}
M.~T.~M. Woodley, L.~Hill, L.~Del~Bino, \emph{et~al.}, \enquote{Self-switching kerr oscillations of counterpropagating light in microresonators,} {\protect\JournalTitle{Phys. Rev. Lett.}} \textbf{126}, 043901 (2021).

\bibitem{delbino2018microresonator}
L.~D. Bino, J.~M. Silver, M.~T.~M. Woodley, \emph{et~al.}, \enquote{Microresonator isolators and circulators based on the intrinsic nonreciprocity of the kerr effect,} {\protect\JournalTitle{Optica}} \textbf{5}, 279--282 (2018).

\bibitem{hill2024symmetry}
L.~Hill, E.-M. Hirmer, G.~Campbell, \emph{et~al.}, \enquote{Symmetry broken vectorial kerr frequency combs from fabry-p{\'e}rot resonators,} {\protect\JournalTitle{Communications Physics}} \textbf{7}, 82 (2024).

\bibitem{del2007optical}
P.~Del’Haye, A.~Schliesser, O.~Arcizet, \emph{et~al.}, \enquote{Optical frequency comb generation from a monolithic microresonator,} {\protect\JournalTitle{Nature}} \textbf{450}, 1214--1217 (2007).

\bibitem{trocha2018ultrafast}
P.~Trocha, M.~Karpov, D.~Ganin, \emph{et~al.}, \enquote{Ultrafast optical ranging using microresonator soliton frequency combs,} {\protect\JournalTitle{Science}} \textbf{359}, 887--891 (2018).

\bibitem{yan2024real}
H.~Yan, A.~Ghosh, A.~Pal, \emph{et~al.}, \enquote{Real-time imaging of standing-wave patterns in microresonators,} {\protect\JournalTitle{Proceedings of the National Academy of Sciences}} \textbf{121}, e2313981121 (2024).

\bibitem{yu2017microresonator}
M.~Yu, Y.~Okawachi, A.~G. Griffith, \emph{et~al.}, \enquote{Microresonator-based high-resolution gas spectroscopy,} {\protect\JournalTitle{Optics letters}} \textbf{42}, 4442--4445 (2017).

\bibitem{alic2014frequency}
N.~Alic, \enquote{Frequency combs in telecommunications applications,} in \emph{Frontiers in Optics,}  (Optica Publishing Group, 2014), pp. FM3C--6.

\bibitem{zhang2019terahertz}
S.~Zhang, J.~M. Silver, X.~Shang, \emph{et~al.}, \enquote{Terahertz wave generation using a soliton microcomb,} {\protect\JournalTitle{Optics express}} \textbf{27}, 35257--35266 (2019).

\bibitem{delbino2021optical}
L.~D. Bino, N.~Moroney, and P.~Del'Haye, \enquote{Optical memories and switching dynamics of counterpropagating light states in microresonators,} {\protect\JournalTitle{Opt. Express}} \textbf{29}, 2193--2203 (2021).

\bibitem{white2023integrated}
A.~D. White, G.~H. Ahn, K.~V. Gasse, \emph{et~al.}, \enquote{Integrated passive nonlinear optical isolators,} {\protect\JournalTitle{Nature Photonics}} \textbf{17}, 143--149 (2023).

\bibitem{moroney2020logic}
N.~Moroney, L.~D. Bino, M.~T.~M. Woodley, \emph{et~al.}, \enquote{Logic gates based on interaction of counterpropagating light in microresonators,} {\protect\JournalTitle{J. Lightwave Technol.}} \textbf{38}, 1414--1419 (2020).

\bibitem{melloni2003linear}
A.~Melloni, F.~Morichetti, and M.~Martinelli, \enquote{Linear and nonlinear pulse propagation in coupled resonator slow-wave optical structures,} {\protect\JournalTitle{Optical and Quantum Electronics}} \textbf{35}, 365--379 (2003).

\bibitem{chen2004nonlinearity}
Y.~Chen and S.~Blair, \enquote{Nonlinearity enhancement in finite coupled-resonator slow-light waveguides,} {\protect\JournalTitle{Optics Express}} \textbf{12}, 3353--3366 (2004).

\bibitem{guo2021transition}
S.-T. Guo, Y.-H. Zhang, L.-L. Wu, \emph{et~al.}, \enquote{Transition between coupled-resonator-induced transparency and absorption,} {\protect\JournalTitle{Physical Review A}} \textbf{103}, 033510 (2021).

\bibitem{grudinin2010phonon}
I.~S. Grudinin, H.~Lee, O.~Painter, and K.~J. Vahala, \enquote{Phonon laser action in a tunable two-level system,} {\protect\JournalTitle{Physical review letters}} \textbf{104}, 083901 (2010).

\bibitem{hua2017high}
Q.~Hua, C.~Yang, X.~Jiang, and M.~Xiao, \enquote{High-order filters based on three high-q microtoroid cavities,} in \emph{2017 Opto-Electronics and Communications Conference (OECC) and Photonics Global Conference (PGC),}  (IEEE, 2017), pp. 1--2.

\bibitem{fan2017tunable}
H.~Fan, L.~Fan, C.~Xia, \emph{et~al.}, \enquote{Tunable fano-like resonance analysis based on a system consisting of a two-silica-microdisk-coupled mach--zehnder interferometer and graphene,} {\protect\JournalTitle{JOSA B}} \textbf{34}, 2429--2435 (2017).

\bibitem{wen2014pt}
J.~Wen, L.~Jiang, X.~Jiang, and M.~Xiao, \enquote{Pt-symmetry and on-chip optical nonreciprocity in active-passive-coupled microtoroids,} in \emph{Frontiers in Optics,}  (Optica Publishing Group, 2014), pp. FTh2B--6.

\bibitem{ghosh2023four}
A.~Ghosh, L.~Hill, G.-L. Oppo, and P.~Del'Haye, \enquote{Four-field symmetry breakings in twin-resonator photonic isomers,} {\protect\JournalTitle{Physical Review Research}} \textbf{5}, L042012 (2023).

\bibitem{ghosh2024controlled}
A.~Ghosh, A.~Pal, L.~Hill, \emph{et~al.}, \enquote{Controlled light distribution with coupled microresonator chains via kerr symmetry breaking,} {\protect\JournalTitle{arXiv preprint arXiv:2402.10673}}  (2024).

\bibitem{mai2022nonreciprocal}
J.~Mai and K.~W. Cheah, \enquote{Nonreciprocal transmission in a nonlinear coupled heterostructure,} {\protect\JournalTitle{Optics Express}} \textbf{30}, 46357--46365 (2022).

\bibitem{cheah2023spontaneous}
J.~Mai, X.~Huang, X.~Guo, \emph{et~al.}, \enquote{Spontaneous symmetry breaking of coupled fabry--p{\'e}rot nanocavities,} {\protect\JournalTitle{Communications Physics}} \textbf{7}, 223 (2024).

\bibitem{pidgaiko2023dispersion}
D.~Pidgaiko, J.~Riemensberger, A.~Tusnin, \emph{et~al.}, \enquote{Dispersion engineering with coupled microresonators for extended soliton microcomb control,} in \emph{2023 Conference on Lasers and Electro-Optics (CLEO),}  (IEEE, 2023), pp. 1--2.

\bibitem{miller2015tunable}
S.~A. Miller, Y.~Okawachi, S.~Ramelow, \emph{et~al.}, \enquote{Tunable frequency combs based on dual microring resonators,} {\protect\JournalTitle{Optics express}} \textbf{23}, 21527--21540 (2015).

\bibitem{komagata2021dissipative}
K.~Komagata, A.~Tusnin, J.~Riemensberger, \emph{et~al.}, \enquote{Dissipative kerr solitons in a photonic dimer on both sides of exceptional point,} {\protect\JournalTitle{Communications Physics}} \textbf{4}, 159 (2021).

\bibitem{tikan2021emergent}
A.~Tikan, J.~Riemensberger, K.~Komagata, \emph{et~al.}, \enquote{Emergent nonlinear phenomena in a driven dissipative photonic dimer,} {\protect\JournalTitle{Nature Physics}} \textbf{17}, 604--610 (2021).

\bibitem{sakaguchi2011symmetry}
H.~Sakaguchi and B.~A. Malomed, \enquote{Symmetry breaking of solitons in two-component gross-pitaevskii equations,} {\protect\JournalTitle{Phys. Rev. E}} \textbf{83}, 036608 (2011).

\bibitem{sakaguchi2021symmetry}
H.~Sakaguchi and B.~A. Malomed, \enquote{Symmetry breaking in a two-component system with repulsive interactions and linear coupling,} {\protect\JournalTitle{Communications in Nonlinear Science and Numerical Simulation}} \textbf{92}, 105496 (2021).

\bibitem{del2013laser}
P.~Del'Haye, S.~A. Diddams, and S.~B. Papp, \enquote{Laser-machined ultra-high-q microrod resonators for nonlinear optics,} {\protect\JournalTitle{Applied Physics Letters}} \textbf{102} (2013).

\bibitem{carmon2004dynamical}
T.~Carmon, L.~Yang, and K.~J. Vahala, \enquote{Dynamical thermal behavior and thermal self-stability of microcavities,} {\protect\JournalTitle{Optics express}} \textbf{12}, 4742--4750 (2004).

\bibitem{garbin2020asymmetric}
B.~Garbin, J.~Fatome, G.-L. Oppo, \emph{et~al.}, \enquote{Asymmetric balance in symmetry breaking,} {\protect\JournalTitle{Physical Review Research}} \textbf{2}, 023244 (2020).

\bibitem{pal2023machine}
A.~Pal, A.~Ghosh, S.~Zhang, \emph{et~al.}, \enquote{Machine learning assisted inverse design of microresonators,} {\protect\JournalTitle{Optics Express}} \textbf{31}, 8020--8028 (2023).

\bibitem{helgason2021dissipative}
{\'O}.~B. Helgason, F.~R. Arteaga-Sierra, Z.~Ye, \emph{et~al.}, \enquote{Dissipative solitons in photonic molecules,} {\protect\JournalTitle{Nature Photonics}} \textbf{15}, 305--310 (2021).

\bibitem{yuan2023soliton}
Z.~Yuan, M.~Gao, Y.~Yu, \emph{et~al.}, \enquote{Soliton pulse pairs at multiple colours in normal dispersion microresonators,} {\protect\JournalTitle{Nature Photonics}} \textbf{17}, 977--983 (2023).

\bibitem{zhang2019electronically}
M.~Zhang, C.~Wang, Y.~Hu, \emph{et~al.}, \enquote{Electronically programmable photonic molecule,} {\protect\JournalTitle{Nature Photonics}} \textbf{13}, 36--40 (2019).

\bibitem{peng2014loss}
B.~Peng, {\c{S}}.~{\"O}zdemir, S.~Rotter, \emph{et~al.}, \enquote{Loss-induced suppression and revival of lasing,} {\protect\JournalTitle{Science}} \textbf{346}, 328--332 (2014).

\end{thebibliography}






\end{document}